\documentclass{article}
\usepackage[labelfont=bf]{caption} 
\usepackage[font={small}]{caption}   

\usepackage{graphicx}
\usepackage{dcolumn}
\usepackage{bm}
\usepackage[pdftex,colorlinks=true,linkcolor=blue,citecolor=blue]{hyperref}
\usepackage[utf8]{inputenc}
\usepackage{lipsum}  

\usepackage{amsmath}
\usepackage{authblk}
\usepackage{amsfonts}
\usepackage{placeins}

\usepackage[numbers, sort&compress]{natbib}  
\usepackage{bibunits}
\usepackage{amssymb}
\usepackage{pdflscape} 
\usepackage{multirow}

\usepackage[table, usenames, dvipsnames]{xcolor}
\usepackage{makecell}
\usepackage{boldline}
\setcellgapes{3pt}
\usepackage{amsmath}
\usepackage{amssymb}

\usepackage{textgreek}

\usepackage[strict]{changepage} 
\usepackage{nicefrac}

\defaultbibliographystyle{naturemag}
\defaultbibliography{references}

\usepackage{lineno}
\usepackage{comment}
\usepackage{microtype}

\usepackage{paralist}

\usepackage{xspace}

\title{A micromechanical frequency reference with parts-per-trillion holdover stability}
\author{
    Jie Yan$^{1}$, Jintark Kim$^2$, Rakibul Islam$^1$, Jiawei Yang$^3$, Karim Elmeligy$^{1}$,
    Alkim~Bozkurt$^{1}$, Thomas W. Kenny$^{3}$, Pavan K. Hanumolu$^{1}$, and Gaurav Bahl$^{2*}$\\
    $^1$ Department of Electrical $\&$ Computer Engineering, \\
    $^2$ Department of Mechanical Science $\&$ Engineering, \\
    University of Illinois at Urbana–Champaign, Urbana, IL 61801 USA \\
    $^3$ Department of Mechanical Engineering, \\
    Stanford University, Stanford, CA 94305 USA \\
    $^*$ bahl@illinois.edu\\
}

\date{}

\begin{document}
\begin{bibunit}

\maketitle

\vspace{-24pt}


\begin{abstract}

Microelectromechanical (MEMS) resonators are widely used in timekeeping applications, and recent advances in fabrication, materials, and encapsulation technology have advanced their potential as high stability frequency references. However, for holdover applications that require the highest levels of long-term frequency stability, compact vapor atomic clocks remain dominant.
In this work, we demonstrate a 268 MHz MEMS clock that achieves record fractional frequency stability of $\sim$8 parts-per-trillion at an averaging time of 8 hours, competitive with chip-scale atomic clocks. We achieved this using a single-crystal silicon electrostatic resonator that has no currently known intrinsic drift mechanism and is protected from the environment with a wafer-level encapsulation. 
We specifically identify gain variations in the sustaining electronics as the dominant limitation in conventional phase-locked oscillator architectures -- originating from temperature sensitivity and drifts in the electronic components -- and overcome this by implementing a frequency-locked loop architecture based on dual-frequency resonance tracking (DFRT). This novel approach removes the specific gain of the supporting electronics as a frequency determining variable in the oscillator. When combined with dual-mode tracking and ratiometric temperature stabilization of the resonator, this approach enables a dramatic enhancement to long-term frequency stability and establishes gain-insensitive DFRT locking as a general paradigm for high-stability MEMS clocks. \\

\end{abstract}

Timekeeping underpins modern communication, navigation, sensing and distributed electronic systems. Over the past two decades, clocks based on microelectromechanical (MEMS) resonators have emerged as promising timing references \cite{nguyen2007mems, van2011review, wu2020mems, ortiz2020low,wei2024mems, kim2025ultra, yang2025precision},
offering an alternative to temperature stabilized quartz references and compact vapor atomic clocks (e.g. chip-scale atomic clocks, or CSACs).  
Unlike low phase-noise oscillators, 
that exhibit very low short-term fluctuations or jitter, frequency references are designed for extreme stability over long durations.
This frequency stability, measured as a fractional frequency fluctuation, is governed by distinct mechanisms across timescales and is a key figure of merit for all timing references \cite{rubiola2008phase}. 
In MEMS-based clocks, short-term stability is primarily limited by the resonator quality factor and signal-to-noise ratio \cite{leeson1966simple}, while long-term stability is constrained by intrinsic aging and systemic drifts. 
At intermediate timescales of hours to days, however, temperature-induced frequency shift becomes the dominant source of instability due to temperature variations in both the resonator and the supporting electronics.
The intermediate and long-duration regimes are particularly critical for holdover operation, e.g. in GPS-denied environments and distributed systems without continuous external calibration.

To address temperature-induced frequency shift, extensive work has been devoted to suppressing the temperature sensitivity of the frequency setting element, i.e., the MEMS resonator. Specifically, variations in ambient temperature alter the mechanics of the resonator (primarily Young’s modulus, dimensions, and thermally induced stress), which in turn can shift the intrinsic resonant frequency of the clock \cite{ng2014temperature, jaakkola2014determination}.
Passive approaches, including degenerate doping engineering \cite{samarao2011temperature, hajjam2012doping}, composite material structures \cite{melamud2009temperature}, crystal-orientation optimization  \cite{shahmohammadi2013turnover, ng2014temperature},
electrode gap tuning \cite{hsu2002stiffness,liu2020temperature}, and stress engineering \cite{perello2024geometrically} have been explored to reduce this intrinsic temperature sensitivity.
Active techniques, such as ovenization, dual-mode temperature stabilization \cite{jha2007high, salvia2009real, ortiz2020low, kim2025ultra, yang2025precision, dabas2025temperature},
and temperature compensation (implemented through electronic or digital correction) \cite{roshan201611, dabas2025temperature, kim2026non},
either maintain the resonator at a fixed temperature or directly counteract the resulting frequency shift.

\begin{figure}[tb]
    \begin{adjustwidth*}{-1in}{-1in}
    \hsize=\linewidth
    \includegraphics[width=1.4\columnwidth]{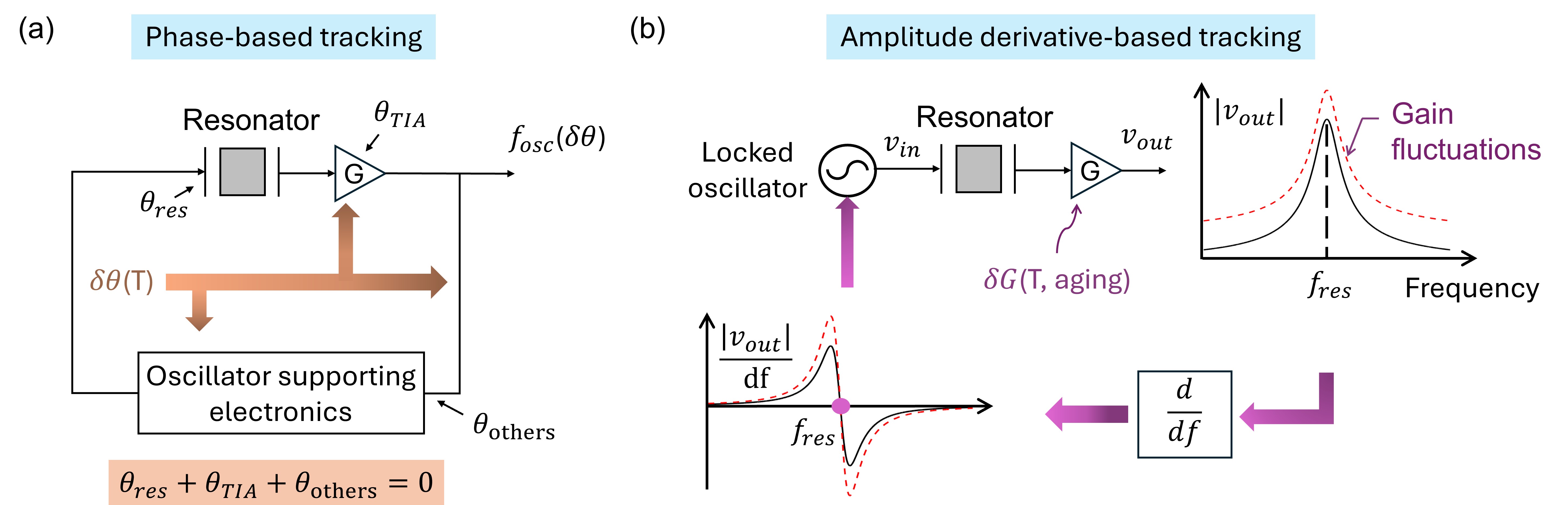}
    \centering
    \caption{
        \textbf{Comparison of frequency tracking architectures.}
        \textbf{(a)} 
        Phase-based tracking, used in self-sustaining oscillators and phase-locked loops, relies on a fixed loop phase condition to set the oscillation frequency. This approach is sensitive to variations in the electronics contribution to the loop phase $\delta\theta(T)$, which shift the locking point and introduce frequency error.
        \textbf{(b)} 
        Frequency tracking based on amplitude derivatives does not rely on loop phase information. The lock point is defined by the symmetry of the magnitude response, rendering the system insensitive to both loop phase variations and gain fluctuations.
    }
    \label{fig1}
    \end{adjustwidth*}
\end{figure}

Even when the resonator itself is nominally stabilized, however, MEMS clocks still exhibit residual frequency fluctuations correlated with ambient temperature because the temperature sensitivity of the sustaining electronics is typically not compensated \cite{salvia2009real, ortiz2020low, kim2025ultra, yan2025electronics}. 
In particular, temperature variations alter the gain (both amplitude and phase responses) of electronic components such as readout amplifiers and phase detectors, introducing phase shift in the feedback loop. Because resonator-based clocks operate either as self-sustaining oscillators or within phase-locked loop (PLL) architectures where the oscillation frequency is defined by a fixed loop-phase condition (typically a zero-phase setpoint), these phase variations shift the loop locking point and translate into frequency error (Fig.~\ref{fig1}(a)). Consequently, temperature-dependent phase shift in the sustaining electronics emerges as a dominant limitation in current MEMS clock architectures. This issue also applies to aging related long-term drifts in the supporting electronics.

This motivates an alternative architecture that removes phase from the locking condition and thereby prevents electronics phase shift from mapping onto frequency error. Instead of enforcing a fixed loop-phase condition, the resonant frequency may be determined from the peak of the amplitude response. As illustrated in Fig.~\ref{fig1}(b), locking to the amplitude peak, corresponding to the zero crossing of the amplitude derivative with respect to frequency, produces a frequency discriminator that is inherently insensitive to electronic phase delay and gain variations. 
A practical way to approximate this amplitude derivative is through a two-point measurement scheme. Dual-frequency resonance tracking (DFRT), originally developed for atomic force microscopy \cite{rodriguez2007dual}, probes the resonator at two closely spaced frequencies and forms an error signal from the difference in their measured amplitudes. 
Conceptually similar frequency discriminators are widely used in atomic clocks as well, where square-wave frequency modulation probes the atomic transition at two frequencies and the resulting difference in transition probability generates the feedback error signal \cite{sullivan2001primary, martinez2023chip}.

Here, we demonstrate an ultra-stable MEMS clock that is insensitive to temperature variations and exhibits a fractional frequency stability of \mbox{$8\times 10^{-12}$} at an averaging time of 8~hours. 
We achieve this performance using a single-crystal silicon MEMS resonator that is hermetically encapsulated at the wafer level and passively temperature-compensated through doping. DFRT-based frequency tracking is applied on two modes of this resonator and dual-mode ratiometric thermometry is leveraged for temperature control.
This architecture simultaneously ovenizes the resonator at a fixed temperature while mitigating sensitivity to drifts in the gain of the supporting electronics.

\section*{Operating Principle and System Architecture}

\begin{figure}[tb]
    \begin{adjustwidth*}{-1in}{-1in}
    \hsize=\linewidth
    \includegraphics[width=1.3\columnwidth]{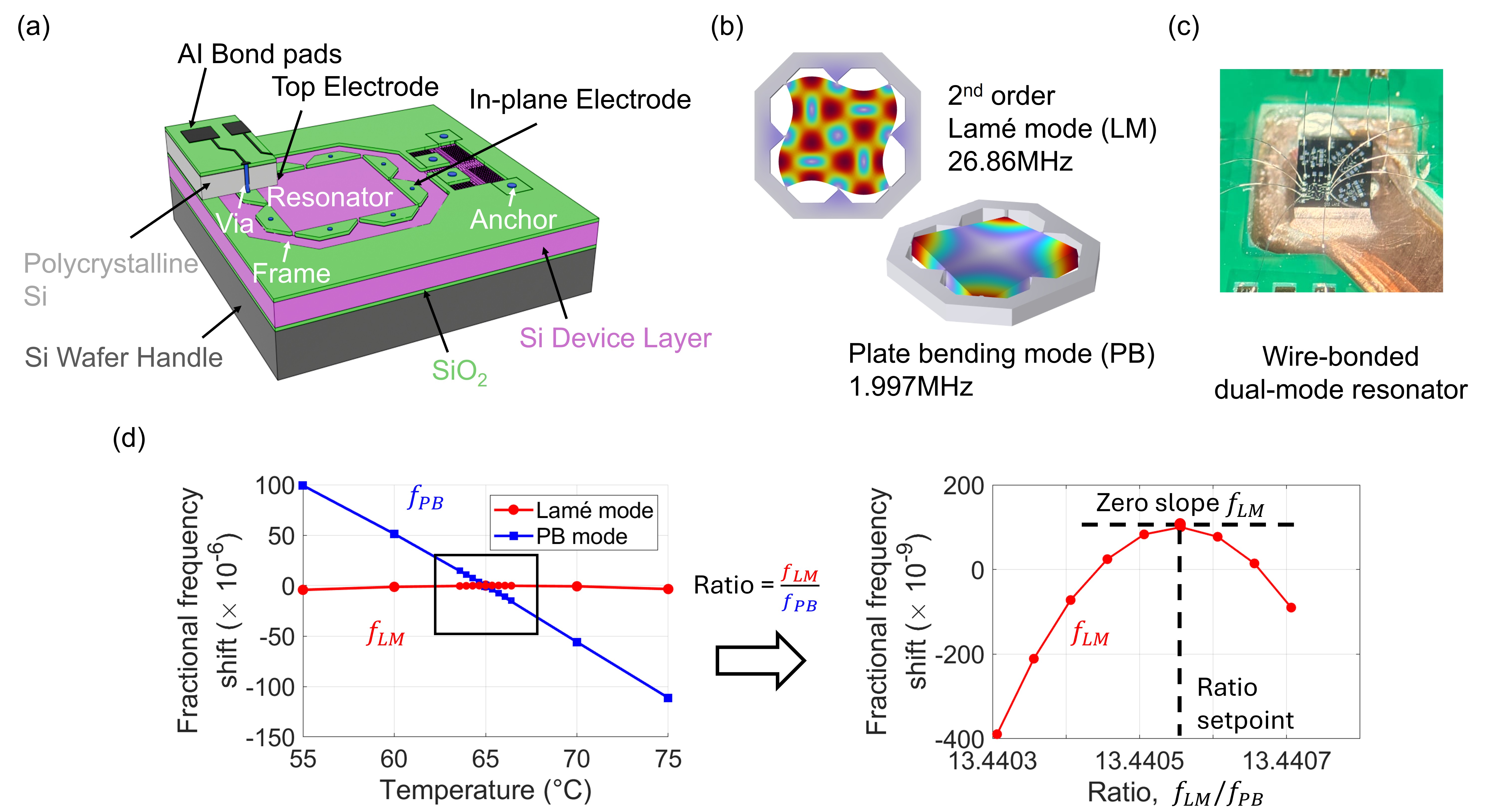}
    \centering
    \caption{
        \textbf{Dual-mode wafer level encapsulated resonator and its temperature characteristics.}
        \textbf{(a)} 
        Three-dimensional rendering of the silicon square plate resonator fabricated using a wafer level encapsulation process. 
        \textbf{(b)} 
        The resonator body is anchored in the middle of each edge, and supports two primary vibration modes: a plate-bending mode and a second-order Lam\'e mode. Each mode is driven and sensed differentially.
        \textbf{(c)} 
        Photograph of a resonator die mounted on a cupper chuck and wire-bonded to an interface printed circuit board. 
        \textbf{(d)} 
        Experimentally measured deviation of the resonance frequencies of both modes as a function of temperature. Data are acquired near the Lam\'e-mode turnover point $\sim 65$ $^{\circ}$C. The boxed region is re-expressed in terms of the frequency ratio, which provides temperature information for active thermal stabilization.
    }
    \label{fig2}
    \end{adjustwidth*}
\end{figure}

The reference used in this clock (Fig.~\ref{fig2}(a)) is an electrostatic MEMS square plate resonator, previously reported in \cite{ortiz2020low}. 
The resonator body is composed entirely of single crystal silicon, $300\times300~\mu\mathrm{m}^2$ in area and 40~$\mu$m thick. 
The material is heavily boron-doped ($\sim10^{20}~\mathrm{cm^{-3}}$) to reduce the temperature coefficient of frequency (TCF)~\cite{ng2014temperature}.
The device is suspended using a high thermal-resistance structure and frame, and is anchored to the substrate on only one side to mitigate stress-induced frequency tuning effects. 
A bias voltage is applied through the suspension structure while the adjacent drive and sense electrodes are kept at nominal ground during operation.
The in-plane transduction gap is $\sim700$ nm, while the out-of-plane transduction gap is $\sim 2$ $\mu$m.
The device was fabricated using a wafer-level encapsulation process based on epitaxial silicon sealing~\cite{ng2014etch}, and resides in a hermetically sealed low-pressure hydrogen environment that protects it from external contaminants and from variations in ambient pressure. The final epitaxial sealing step is performed at $\sim$~1100~$^\circ$C, causing any stray contaminants to be removed or to be gettered within the silicon. 
Top-electrode isolation is achieved by etching trenches through the cap and refilling them with silicon nitride.
These devices have no currently known intrinsic drift mechanism \cite{ng2014etch, kim2007frequency}.

The resonator body supports two differentially driven and sensed vibration modes (Fig.~\ref{fig2}(b)): a second-order Lam\'e mode at $26.84~\mathrm{MHz}$ ($Q \approx 6.8\times10^{5}$) and a plate-bending (PB) mode at $1.20~\mathrm{MHz}$ ($Q \approx 9.3\times10^{5}$). The Lam\'e mode exhibits a second-order TCF with a turnover near $65^{\circ}\mathrm{C}$, characterized by fractional coefficients $\alpha_{1}=4.98\times10^{-6}\,\mathrm{^{\circ}C^{-1}}$ and $\alpha_{2}=-0.038~\times10^{-6}\,\mathrm{^{\circ}C^{-2}}$ with the relation $\delta\!f/f = \alpha_1 \,\Delta T + \alpha_2 \,\Delta T^2$, where $\Delta T$ is defined relative to the turnover temperature. 
Stabilizing the device temperature near this turnover point suppresses first-order temperature sensitivity, making the mode very well suited for clock applications.
In contrast, the PB mode exhibits an approximately linear TCF curve with 
$\alpha_1 = -10.5~\times10^{-6}\,\mathrm{^{\circ}C^{-1}}$, making it suitable as an in-situ temperature sensor.
The temperature sensing function can be accomplished by performing a relative count, e.g. using a frequency counter, to determine the frequency ratio between the modes \cite{jha2007high}. The ratio provides a one-to-one mapping to temperature (Fig.~\ref{fig2}(d)) due to the distinct temperature dependencies of the two modes, and can be leveraged for temperature stabilization via feedback control~\cite{salvia2009real, ortiz2020low}. Specifically, locking the frequency ratio to a fixed value (by controlling a heater) effectively stabilizes the device temperature and constrains the Lam\'e-mode frequency to a stable operating point. The ideal setpoint is at the frequency ratio corresponding to the Lam\'e mode turnover, where $f_{LM}$ exhibits no variation.

\begin{figure}[tb]
    \begin{adjustwidth*}{-1in}{-1in}
    \hsize=\linewidth
    \includegraphics[width=1.3\columnwidth]{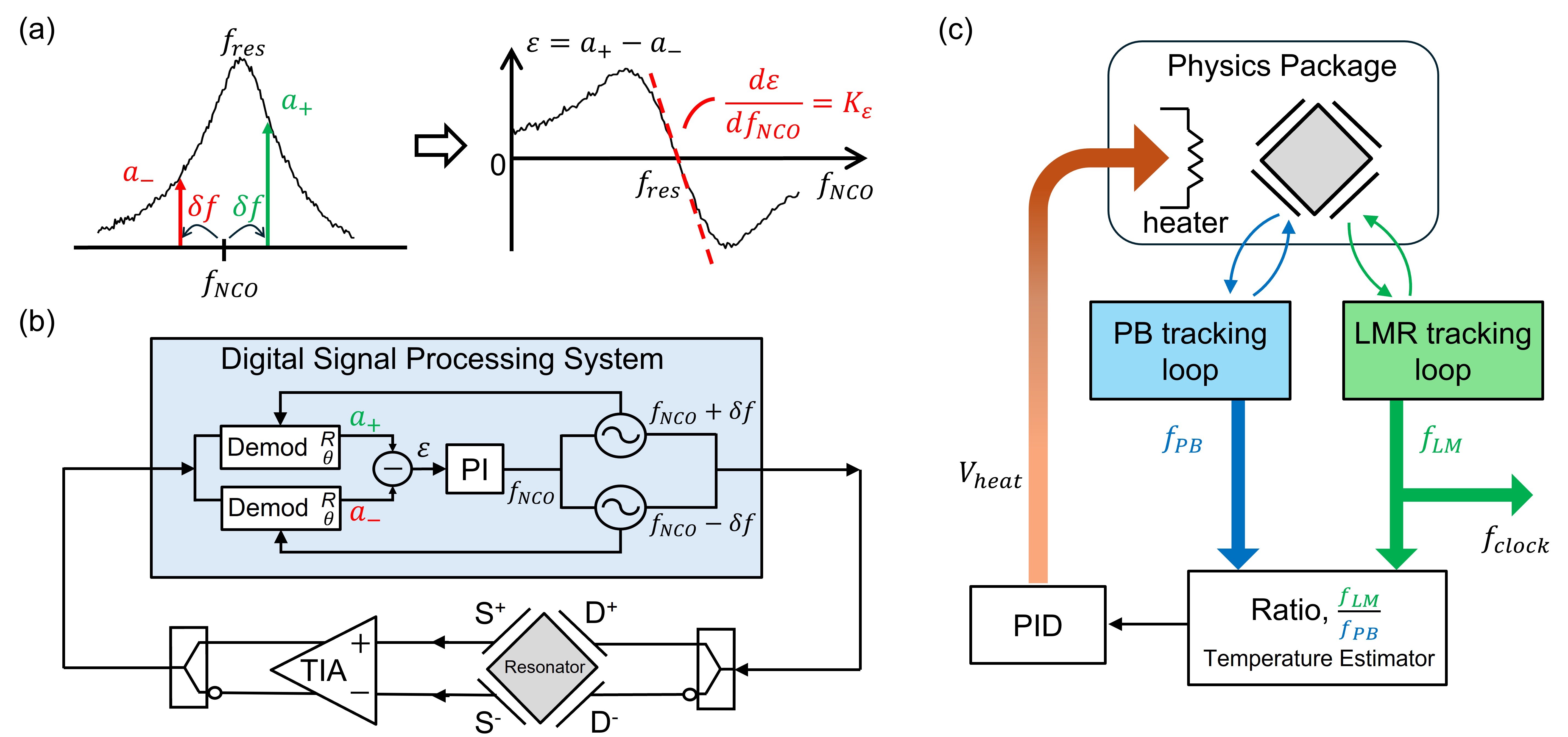}
    \centering
    \caption{
        \textbf{Dual-frequency resonance-tracking (DFRT) principle and clock architecture.}
        \textbf{(a)} A representative resonator magnitude response and the corresponding DFRT error signal. The goal is to set a numerically controlled oscillator $f_{NCO}$ to the peak of the resonance at $f_{res}$. We measure the amplitude responses at $f_\textrm{NCO} \pm \delta\!f$ and evaluate their difference to produce the frequency-tracking error signal $\varepsilon$.
        \textbf{(b)} Architecture of our DFRT-based frequency-locked loop (FLL). A digital signal processing (DSP) unit generates the dual NCO tones at $f_\mathrm{NCO}\pm \delta\!f$. The resonator is excited differentially at both tones simultaneously, and the output motional current is sensed by a differential transimpedance amplifier (TIA). The amplitude responses are individually demodulated (phase information is discarded) and the resulting error signal is processed by a proportional–integral (PI) controller to update $f_{\mathrm{NCO}}$ and close the feedback loop.
        \textbf{(c)} The complete clock implementation combines FLL tracking on both modes, with ratiometric temperature measurement and thermal stabilization.
    }
    \label{fig3}
    \end{adjustwidth*}
\end{figure}

To realize the clock, the resonant frequencies of both modes must be continuously tracked. We accomplished this using a DFRT-based frequency-locked loop (FLL) for each mode, as described earlier. As illustrated in Fig.~\ref{fig3}(a), we probe each resonator mode with two tones injected symmetrically about the nominal frequency $f_\mathrm{NCO}$ with offsets of $\pm\delta\!f$. The difference between the two magnitude responses $a_{\pm}$ for each tone produces a derivative-like error signal that is monotonic near resonance and crosses zero when $f_{\mathrm{NCO}}$ coincides with the resonant peak. This signal therefore functions as a frequency discriminator that directly identifies the maximum of the magnitude response, independent of gain variations in the supporting electronics. 
The offset frequency $\delta\!f$ is selected near the point of maximum amplitude sensitivity to minimize noise-to-frequency conversion. For an ideal Lorentzian response, this occurs at
$\delta\!f = f_0/(2\sqrt{2}Q)$.

The implementation of the FLL architecture for a single mode is shown in Fig.~\ref{fig3}(b). The resonator is driven simultaneously by two numerically controlled oscillator (NCO) tones in the digital signal processing (DSP) system at $f_{\mathrm{NCO}}\pm\delta\!f$. The motional response is digitally demodulated to extract the amplitudes at the two sidebands, $a_{-}$ and $a_{+}$, whose difference is the frequency-tracking error signal, $\varepsilon = a_{+} - a_{-}$. This error is processed by a proportional–integral (PI) controller to update $f_{\mathrm{NCO}}$, which locks to the resonance condition and serves as the clock output.

\section*{Experimental Results}
The complete system architecture is implemented as shown in Fig.~\ref{fig3}(c). The resonator die is mounted on a cylindrical copper chuck wrapped with nichrome heating wire for temperature regulation, and is attached using silver paste and wire-bonded to a printed circuit board (Fig.~\ref{fig2}(c)). The assembly is enclosed in a simple container to form the physics package and minimize airflow effects. 
A 10~V dc bias is applied to the resonator body to enable electrostatic actuation and generate an output motional current, provided by an SRS DC205 voltage source. The differentially driven and sensed motional currents are converted to voltages using a custom transimpedance amplifier (TIA) based on the ADA4817 operational amplifier with 100~k$\Omega$ feedback resistors. The resulting voltage signals are supplied to a Zurich Instruments UHFLI lock-in amplifier, which serves as the DSP platform for demodulation and FLL control.

\begin{figure}[tb]
    \begin{adjustwidth*}{-1in}{-1in}
    \hsize=\linewidth
    \includegraphics[width=1.4\columnwidth]{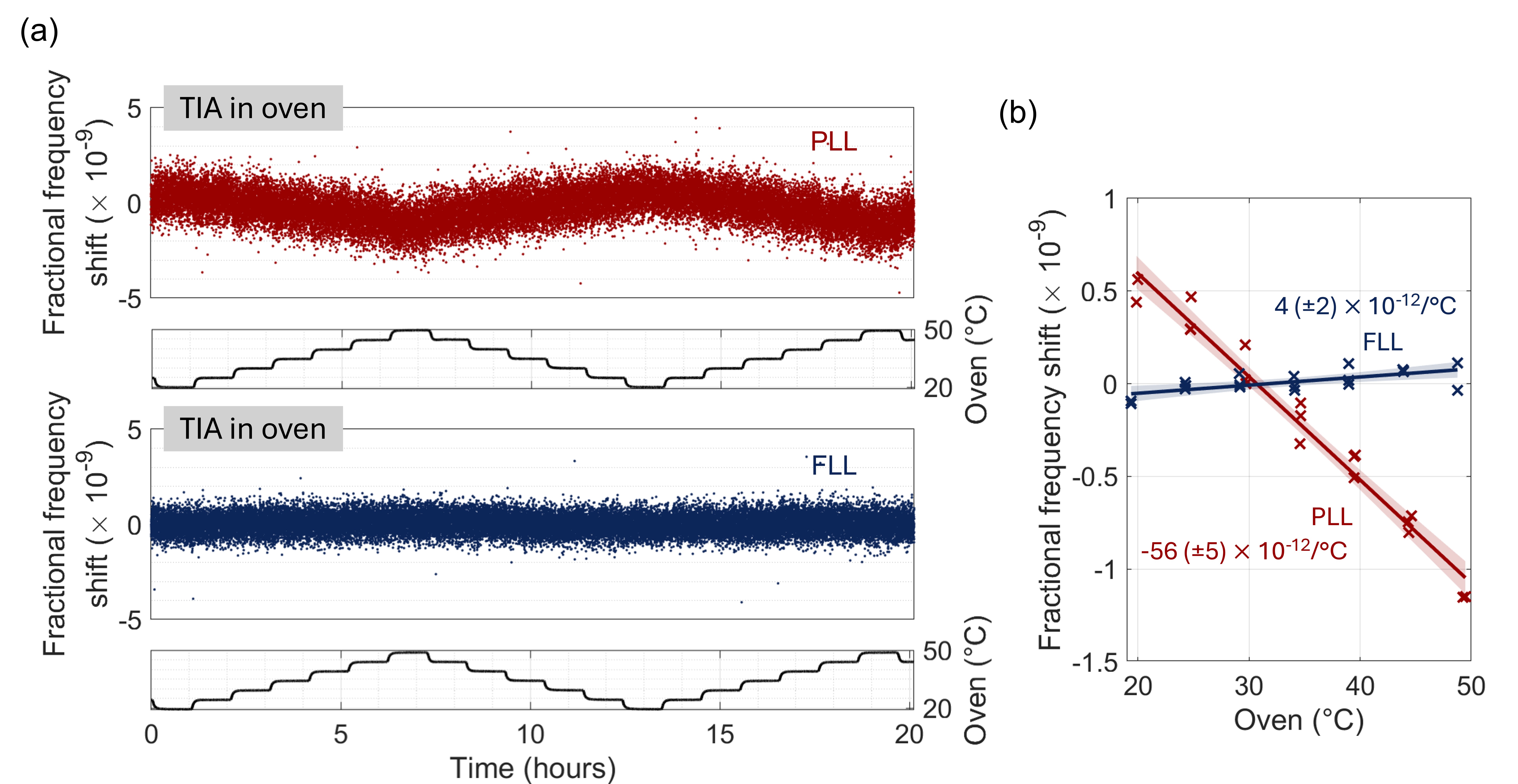}
    \centering
    \caption{
    \textbf{Clock characterization with large temperature deviations applied to the TIAs (20--50~$^{\circ}$C).}        
    \textbf{(a)} We tested the clock to compare both the conventional PLL tracking system (top row) and the DFRT-based FLL system (third row). The large temperature variations are only applied to the TIAs with the goal to affect their transfer functions. The clock output frequency is digitally synthesized as $10 \times$ the Lam\'e resonance frequency, as described in the text. The oven temperature profiles during the experiments are also shown.
    \textbf{(b)} A comparison of the extracted frequency-temperature relationship for both tracking systems tested in (a), showing that the FLL based clock exhibits a temperature coefficient about $14\times$ smaller (within error bars) than that of the PLL version, confirming the superiority of the FLL approach.
    }
    \label{fig4}
    \end{adjustwidth*}
\end{figure}

Each vibration mode is tracked using an independent FLL loop. The Lam\'e mode provides the clock output, while the Lam\'e/PB frequency ratio serves as an intrinsic temperature sensor for active stabilization. Offset frequencies $\delta\!f$ = 14~Hz for the Lam\'e mode and $\delta\!f$ = 3~Hz for the PB mode are selected to balance sensitivity and tracking range. 
The numerical frequencies of both modes from the DSP are used to compute their ratio in real time. As shown previously in Fig.~\ref{fig2}(d), the zero sensitivity (or ``turnover") point of the Lam\'e frequency $f_{LM}$ is observed at ratio = 13.44055. The instantaneous ratio value is processed by a software-implemented PID controller to regulate the voltage applied to the chuck heater.
The tracked Lam\'e-mode frequency is digitally multiplied by ten within the DSP to generate an output near 268~MHz. This output is continuously monitored using a GPS-referenced frequency counter (Tektronix FCA3020) for long-term stability evaluation.

To show that there is reduced sensitivity to changes in loop phase or gain, we intentionally introduced large temperature variations to the measurement electronics. Both TIAs, whose transfer functions are known to be temperature sensitive, were placed inside a temperature controlled oven (TestEquity 115A-F) and the oven temperature was varied between $20$--$50~^{\circ}$C. 
The resonator physics package, splitters, combiners, and Zurich UHFLI remain in room air ($\sim$25~$^{\circ}$C). During the experiment, the clock uses dual-mode stabilization to regulate the resonator temperature. We can now compare two DSP configurations: a conventional PLL and the new DFRT-based FLL.

As shown in Fig.~\ref{fig4}(a), the PLL exhibits a clear temperature-correlated frequency shift with oven temperature variations.. If we assume that this shift only arises from Lam\'e-mode tracking error, we can estimate that the feedback electronics phase varied by $2.4\times10^{-3}$~rad over the temperature cycle (using the relation
$\left.\nicefrac{d\phi}{df}\right|_{f_{0}} \approx \nicefrac{2Q}{f_0}$).
In contrast, the FLL shows almost no frequency variation to the eye in Fig.~\ref{fig4}(a). We extracted the temperature coefficient for both systems in Fig.~\ref{fig4}(b). The FLL exhibits a significant reduction in temperature sensitivity compared to the PLL.

\begin{figure}[tb]
    \begin{adjustwidth*}{-1in}{-1in}
    \hsize=\linewidth
    \includegraphics[width=1.45\columnwidth]{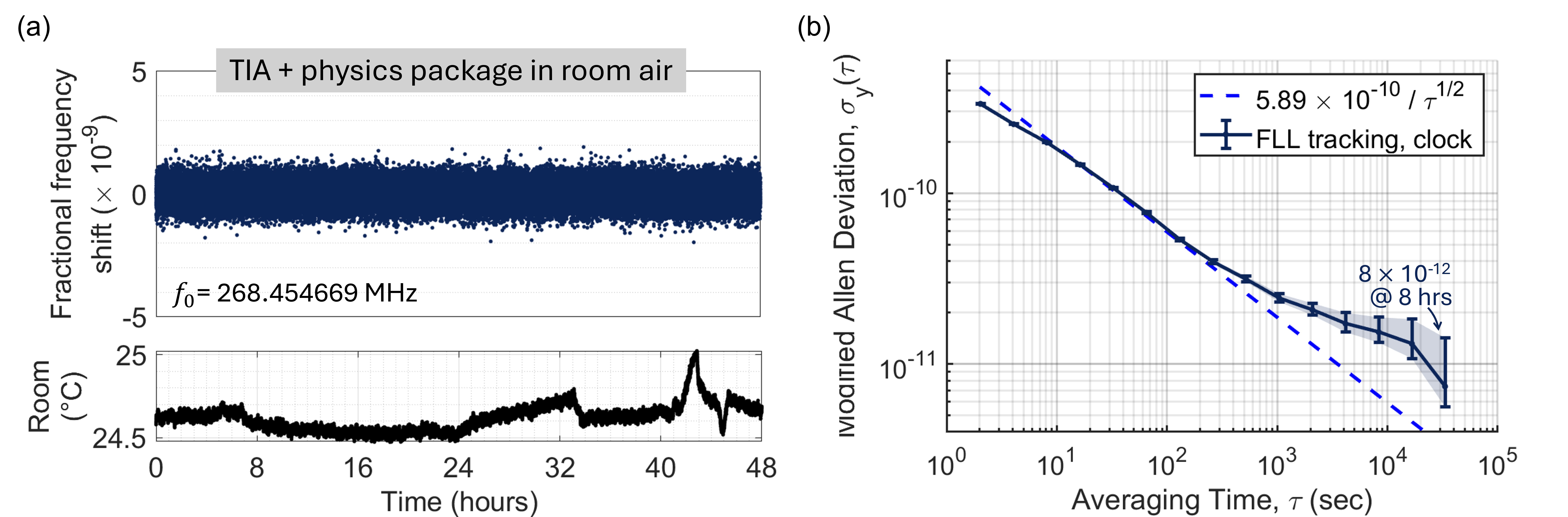}
    \centering
    \caption{
        \textbf{Long-term operation of the 268 MHz clock in ambient laboratory conditions.}
        \textbf{(a)} 
        The fractional frequency record measured over 48 hours of continuous operation is presented along with the ambient temperature record.
        \textbf{(b)} 
        Corresponding modified Allan deviation (MDEV) calculated from the fractional frequency data in (a). The dashed line indicates the predicted white-noise floor based on the measured voltage noise in the DFRT error signal. In Supplementary Fig.~\ref{figs1}, we show that the MDEV limit due to bias voltage fluctuations is much lower than our observations.
    }
    \label{fig5}
    \end{adjustwidth*}
\end{figure}

We next evaluated the long-term FLL clock performance. A continuous 48-hour measurement was conducted with all electronic components exposed to ambient laboratory conditions, while dual-mode stabilization actively regulates the resonator die temperature only.
The measured frequency record is shown in Fig.~\ref{fig5}(a). Frequency measurements were continuously performed with a 1-second gate time, and no discernible correlation with ambient temperature was observed throughout the experiment. A linear fit to the frequency data suggests only the possibility of a slow drift of approximately $-4.1 (\pm 2.5)\times10^{-12}/\mathrm{day}$ with large error bars.

We further analyzed the frequency record via modified Allan deviation (MDEV), shown in Fig.~\ref{fig5}(b), which reaches 8~$\times10^{-12}$ at an averaging time of 8~hours (28,800 seconds). At shorter averaging times, the MDEV follows the expected $\tau^{-1/2}$ scaling characteristic of white frequency noise. For white FM noise, MDEV can be expressed as 
$\mathrm{MDEV}(\tau)=V_{r,n}/(2K_{\varepsilon}f_0\sqrt{\tau})$ \cite{rubiola2008phase, vanier2012signal}.
where $V_{r,n}$ denotes the voltage noise spectral density of the error signal measured far from resonance, $K_{\varepsilon}$ is the zero-crossing slope of the error signal (Fig.~\ref{fig3}(a)), and $f_{0}$ is the resonant frequency.
Using the experimentally measured parameters
($V_{r,n}=5.06\times10^{-6}~\mathrm{V}/\sqrt{\mathrm{Hz}}$,
$K_{\varepsilon}=1.6\times10^{-4}~\mathrm{V/Hz}$, and
$f_{0}=26.84~\mathrm{MHz}$), we obtain a short-timescale predicted 
$\mathrm{MDEV}(\tau)=5.89\times10^{-10}\,\tau^{-1/2}$.
The predicted fluctuation trend closely matches the measured short-term stability.
At longer averaging times, the MDEV departs from the $\tau^{-1/2}$ trend and exhibits a ``bump'' beyond $\sim$4000~seconds. This characteristic is often encountered in clock systems and suggests the presence of residual temperature fluctuations (e.g. due to day-night cycles) or other nonideal effects discussed below.

To the best of our knowledge, our clock achieves the lowest measured 8-hour MDEV of any MEMS-based clock (e.g. \cite{ortiz2020low,wei2024mems, kim2025ultra, yang2025precision, kim2026non}), validated using a dataset that is six times the longest reported averaging duration to ensure statistical reliability. For clocks based on similar resonator designs \cite{ortiz2020low}, this corresponds to an improvement of over two orders of magnitude (approximately $300\times$) compared to prior art. This MDEV result is also comparable to that of compact vapor atomic clocks like CSAC \cite{microchip_sa45s_csac}. State-of-the-art CSACs achieve fractional frequency stabilities in the low $10^{-12}$ range at long averaging times ($10^{4}$–$10^{5}$ seconds) \cite{zhang2016rubidium, rivera2025microcell}, while commercially available CSACs specify a stability of $1 \times 10^{-11}$ at 1,000 seconds, followed by a flicker floor and long-term drift corresponding to an extrapolated stability on the order of $10^{-11}$ at 8 hours \cite{rybak2021chip}. 

\section*{Nonidealities and Practical Limits}
\begin{figure}[htb]
    \begin{adjustwidth*}{-1in}{-1in}
    \hsize=\linewidth
    \includegraphics[width=1.15\columnwidth]{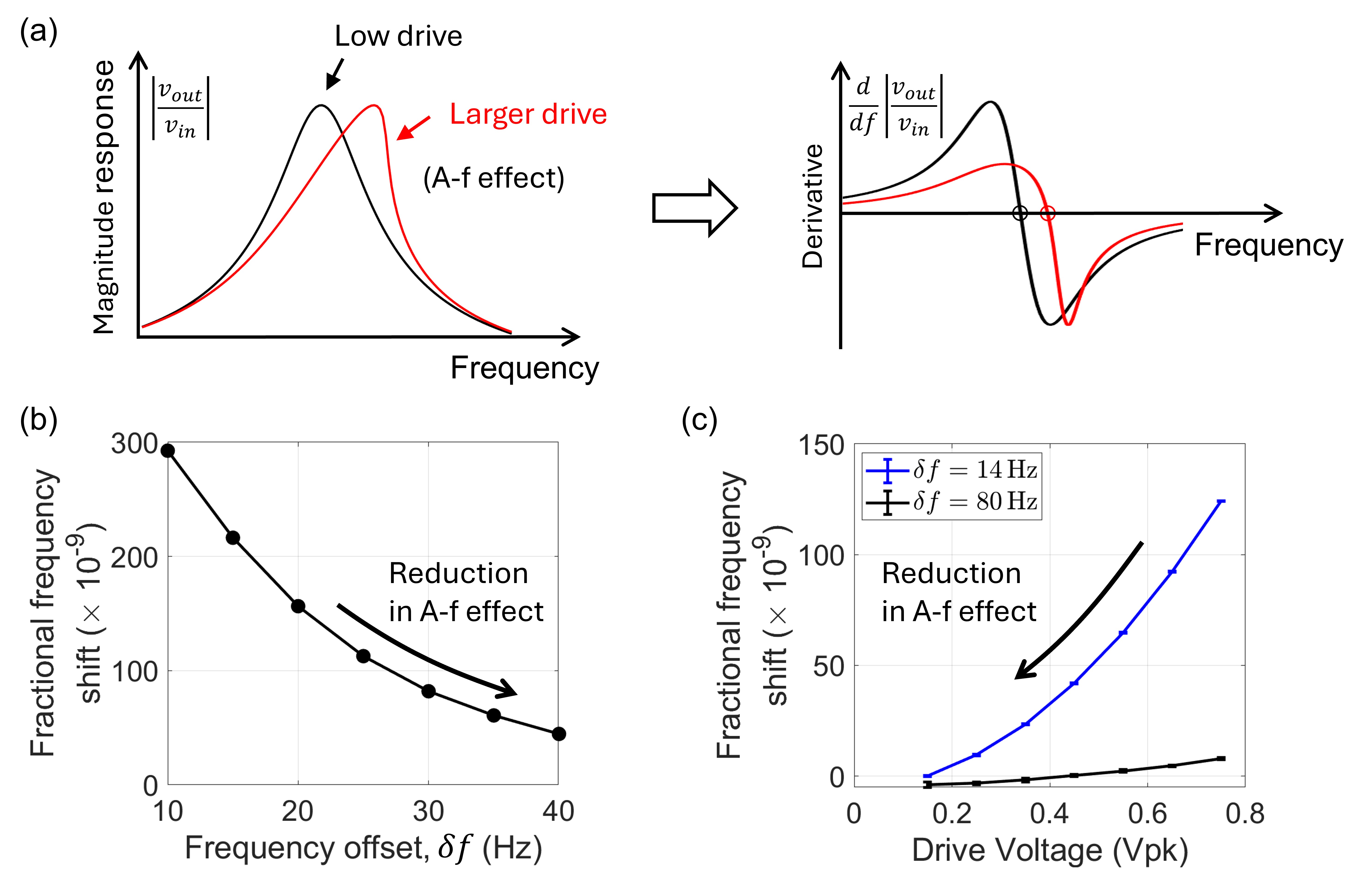}
    \centering
    \caption{
    \textbf{Influence of the amplitude–frequency (A–f) nonlinearity in the FLL based clock.}
    \textbf{(a)} At high drive levels, the A–f nonlinearity affects the symmetry of the resonance magnitude response, shifting the lock point that is obtained through the derivative of the amplitude response.
    \textbf{(b)} Experimental measurements of the clock output frequency as a function of probe tone offset $\delta\!f$, at a fixed drive level. We see that increasing $\delta\!f$ suppresses the observed A–f effect since the drive tones are placed further away from the resonance at $f_{res}$.
    \textbf{(c)} Measurements at $\delta\!f = 14\,\mathrm{Hz}$ and $80\,\mathrm{Hz}$ demonstrate that reducing the drive level also mitigates the A–f effect, as expected. Flat lines correspond to the error bars for each measurement.
    }
    \label{fig6}
    \end{adjustwidth*}
\end{figure}

While the FLL approach, in principle, eliminates sensitivity to variations of loop gain, amplitude-domain nonidealities can still influence frequency stability. Examples include the AC Stark shift in atomic references \cite{sullivan2001primary} and the amplitude–frequency (Duffing) effect in mechanical resonators \cite{lifshitz2008nonlinear}. 
In the two-point FLL scheme, accurate peak localization requires a symmetric magnitude response about resonance. However, at high drive levels, it is well known that amplitude–frequency (A–f) nonlinearity distorts the resonance curve (Fig.~\ref{fig6}(a)), breaking this symmetry. This shifts the effective lock point of the error signal in the FLL scheme. For the Lam\'e bulk-mode resonators used here, the Duffing nonlinearity produces in a positive frequency shift with increasing drive amplitude.

One solution to mitigate the A-f effect is to simply reduce the drive level. In our FLL scheme, it can also be mitigated by pushing the probing tones further away from the true resonance to effectively reduce the motional amplitude, both of which come at the expense of reduced signal-to-noise ratio (SNR). In Fig.~\ref{fig6}(b), we confirm that both A-f induced frequency shift and the frequency sensitivity decrease with increasing offset $\delta\!f$. Fig.~\ref{fig6}(c) further confirms this lower sensitivity to drive level for larger offset $\delta\!f$, while also showing that lower drive level mitigates the A-f effect independent of $\delta\!f$. In our experiments,
when testing the long term performance of the clock (e.g. Fig.~\ref{fig5}), we first set $\delta\!f$ near the optimal value $\delta\!f = f_0/(2\sqrt{2}Q)$ to maximize error signal sensitivity. The drive amplitude was then determined by monitoring the short-term MDEV and increased until nonlinear distortion limits further improvement.

Another potential concern arises from finite demodulation selectivity within the lock-in detection system. 
Under dual-tone excitation, each demodulation channel (which conducts a mixing and low-pass filtering operation on each quadrature) contains a small residual component at $2\,\delta\!f$ due to the adjacent tone. 
This leakage introduces a weak ripple in the generated error signal and constrains the achievable loop bandwidth, since the demodulation filter cutoff frequency must remain well below $2\,\delta\! f$ to sufficiently suppress the mixing product.

Although the finite selectivity introduces a time variation, it does not shift the zero-crossing condition that defines the lock point, as the leakage component is common to both channels and largely cancels during the subtraction performed to get the error signal 
$\varepsilon$. This cancellation can be understood by examining the demodulated amplitudes of the two channels, which can be expressed as
\begin{align}
a_{+\delta\!f}(t) &= \sqrt{\,a_{+}^2 + \alpha\,a_{+}a_{-}\cos\!\big(4\pi \, \delta\!f\, t + \theta_{+}-\theta_{-}\big)\,}, \\
a_{-\delta\!f}(t) &= \sqrt{\,a_{-}^2 + \alpha\,a_{-}a_{+}\cos\!\big(4\pi \, \delta\!f\, t + \theta_{+}-\theta_{-}\big)\,},
\end{align}
where $\alpha$ denotes the residual leakage factor determined by the rejection of the demodulation filter (e.g., $\sim\!60\,\mathrm{dB}$ suppression), $\theta_{+}$ and $\theta_{-}$ represent the phase responses of the two sideband tones, and $a_{+}$ and $a_{-}$ are the true amplitudes of the $+\delta\!f$ and $-\delta\!f$ tones, respectively.
For $|\alpha|\ll 1$, a first-order expansion yields
\begin{align}
a_{+\delta\!f}(t) &\approx a_+ + \frac{\alpha a_-}{2}\cos\!\big(4\pi \, \delta\!f\, t + \theta_{+}-\theta_{-}\big),\\
a_{-\delta\!f}(t) &\approx a_- + \frac{\alpha a_+}{2}\cos\!\big(4\pi \, \delta\!f\, t + \theta_{+}-\theta_{-}\big).
\end{align}
The resulting two-point error signal becomes
\begin{equation}
\varepsilon(t)=a_{+\delta\!f}(t)-a_{-\delta\!f}(t)
\approx (a_+-a_-)
\left[
1-\frac{\alpha}{2}\cos\!\big(4\pi \, \delta\!f\, t + \theta_{+}-\theta_{-}\big)
\right].
\end{equation}
where the time varying term is clearly seen.
Since the controller enforces $\varepsilon(t)=0$, the condition $a_+ = a_-$ must be true at all times. 
Therefore, finite demodulation leakage does not shift the lock point.

As reported above, the measured MDEV is seen to deviate from the ideal $\tau^{-1/2}$ scaling, exhibiting a ``bump'' at long averaging times. This behavior indicates the presence of very low-frequency processes that might arise from a combination of flicker ($1/f$-like) frequency noise and residual temperature sensitivity.
Although the MDEV decreases again at longer $\tau$, the associated uncertainty prevents us from ruling out an underlying flicker noise contribution,
 which would manifest as a sustained plateau over a broad range of $\tau$ \cite{rubiola2008phase}. 
A well-known source of flicker noise in electrostatic resonators is the bias voltage supply, as fluctuations in the applied bias can introduce $1/f$ frequency fluctuations in the resonator \cite{kaajakari2005phase}.
However, with a low-noise bias source, this contribution is expected to be negligible (below the $10^{-13}$ level) and therefore cannot account for the observed behavior (see Supplementary Fig.~\ref{figs1}). 

Residual temperature sensitivity may further arise from drive-level variations induced by temperature fluctuations in the drive source, as A-f nonlinearity can shift the resonator locking point. Slowly varying nonlinear processes may also contribute to this ``bump'' behavior and could result from aging in the drive electronics or long-term changes in interconnects, such as wirebonds or feedthrough components, which modify the parasitic electrical loading of the resonator. Since the extracted linear drift over the full dataset is relatively small $-4.1 (\pm 2.5)\times10^{-12}/\mathrm{day}$, such time-varying aging effects are not captured by this linear estimate. 

\section*{Conclusion}
We have demonstrated a MEMS-based clock achieving a record MDEV of $8 \times 10^{-12}$ at an 8-hour averaging time, which represents the highest reported frequency stability of a micromechanical reference at this timescale.
This performance is enabled by the combined implementation of dual-mode temperature stabilization and an DFRT-based FLL tracking system. The dual-mode scheme suppresses intrinsic resonator temperature sensitivity, while the FLL removes dependence on loop phase, thereby preventing electronics drifts from mapping into frequency error. Controlled oven-sweep experiments were used to confirm strong suppression of electronics temperature sensitivity compared with traditional PLL-type frequency tracking. 
Together, these approaches address the dominant intermediate-timescale instabilities arising from temperature variations in both the resonator and the supporting electronics.

The remaining limitations likely arise primarily from amplitude–frequency nonlinearity, intrinsic resonator mechanisms, and readout noise, highlighting opportunities for further improvement through nonlinear mitigation or compensation, resonator design optimization, and noise reduction. 
Nevertheless, these results establish amplitude-domain resonance tracking as a robust and fundamentally far superior approach to phase-based tracking for high-stability micromechanical timing references. This approach is broadly applicable to a wide class of mechanical resonators for timing, sensing, and navigation, enabling frequency stability that is at par with compact atomic clocks. 

{\footnotesize \nolinenumbers \putbib}

\end{bibunit}

\section*{Acknowledgments}
The authors express their continued gratitude to Dr. Gabrielle Vukasin and Dr. Hyun-Keun Kwon for their work in fabricating the MEMS resonators that were used in these experiments. 
This work was supported by the DARPA H6 program under cooperative agreement HR0011-23-2-0004.
The views and conclusions contained herein are those of the authors and should not be interpreted as necessarily representing the official policies or endorsements, either expressed or implied, of DARPA or the US Government.

\section*{Data availability}

The data that support the findings of this study are available from the corresponding authors upon reasonable request.

\section*{Author contributions}

J.Y. (Jie Yan) and G.B. conceived the work. J.Y. and J.K. performed the experiments, with electronic support from R.I. and A.B.; J.Y. (Jiawei Yang) assisted with device characterization and operation. All authors contributed to data analysis and interpretation. J.Y. (Jie Yan), J.K., and G.B. wrote the manuscript. G.B., T.W.K., and P.K.H. supervised the project.

\FloatBarrier

\newpage

\newcommand{\beginExtFigures}{
        \setcounter{table}{0}
        \renewcommand{\tablename}{Supplementary Data Table}
        \renewcommand{\thetable}{\arabic{table}}
        \setcounter{figure}{0}
        \renewcommand{\figurename}{Supplementary Data Figure}        
        \renewcommand{\thefigure}{\arabic{figure}}
}
\beginExtFigures

\begin{figure}[htb]
    \begin{adjustwidth*}{-1in}{-1in}
    \hsize=\linewidth
    \includegraphics[width=1.3\columnwidth]{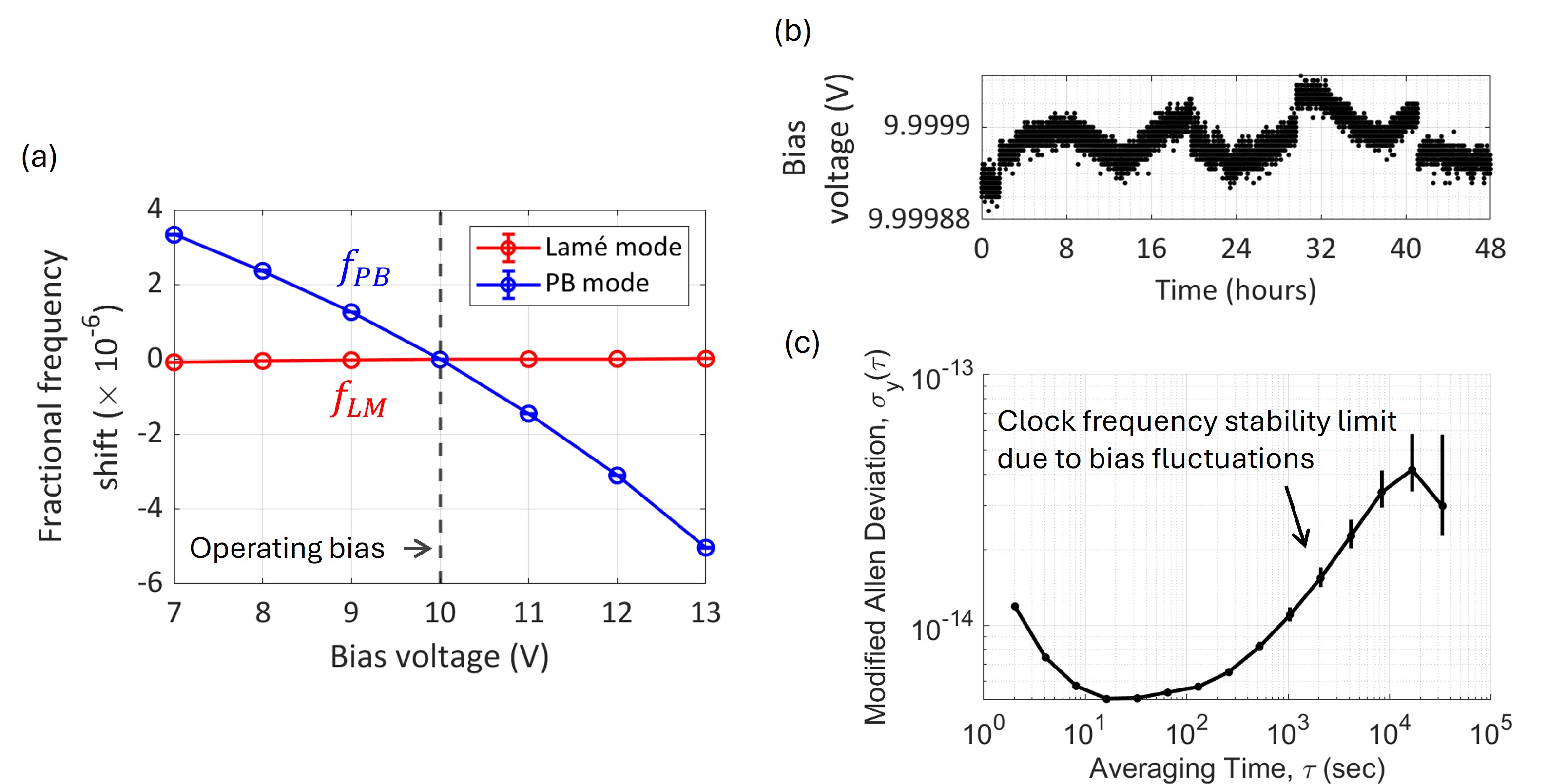}
    \centering
    \caption{
    \textbf{Bias fluctuation induced frequency limits.} The dc bias voltage on the resonator directly changes its resonance frequency through the electrostatic spring softening effect. Subsequently, there are two paths by which the bias can affect the clock frequency: (i) the intrinsic bias sensitivity of the Lamé mode directly changes the clock output, and (ii) the bias-induced frequency shift of the PB mode perturbs the Lamé/PB frequency ratio, causing a fluctuation in the feedback controlled temperature. 
    \textbf{(a)} Independent bias sensitivity characterization of both modes near the operating point of $65^\circ\mathrm{C}$. Linearized at $10\mathrm{V}$ bias, the sensitivities are $-1.35$~ppm/V for the PB mode and $13$~ppb/V for the Lamé mode.
    \textbf{(b)} Measured bias voltage fluctuations using a Fluke 8558A high precision multimeter during the long-term clock experiment in Fig.~\ref{fig5}.
    \textbf{(c)} We derive the lower limit of MDEV from the linearized bias-to-frequency relations in (a) and the bias fluctuation data in (b), accounting for both the intrinsic bias sensitivity of the Lam\'e mode and the PB-mode bias-induced temperature instability. Here we have assumed a very large 0.2~$^{\circ}$C unintentional offset from the true turnover temperature (the real offset is likely closer to 0.01$^{\circ}$C). Our estimate shows that the bias-induced frequency instability is comfortably at or below the $10^{-13}$ level.
    }
    \label{figs1}
    \end{adjustwidth*}
\end{figure}

\end{document}